# Combined stellar structure and atmosphere models for massive stars

## I. Interior evolution and wind properties on the main sequence

D. Schaerer[1], A. de Koter[2], W. Schmutz[3], and A. Maeder[1]

[1] Geneva Observatory, CH-1290 Sauverny, Switzerland; e–mail: schaerer@scsun.unige.ch, maeder@scsun.unige.ch
[2] Universities Space Research Association, Code 681, GSCF, Greenbelt, MD 20771, USA; e–mail: alex@homie.gsfc.nasa.gov
[3] Institut für Astronomie, ETH Zentrum, CH-8092 Zürich, Switzerland; e–mail: schmutz@astro.phys.ethz.ch

**Abstract.** We present the first "combined stellar structure and atmosphere models" (*CoStar*) for massive stars, which consistently treat the entire mass loosing star from the center out to the asymptotic wind velocity. The models use up-to-date input physics and state-of-the-art techniques to model both the stellar interior and the spherically expanding non–LTE atmosphere including line blanketing. Our models thus yield consistent predictions regarding not only the basic stellar parameters, including abundances, but also theoretical spectra along evolutionary tracks. On the same ground they allow us to study the influence of stellar winds on evolutionary models.

In this first paper, we present our method and investigate the wind properties and the interior evolution on the main sequence (MS) at solar metallicity.

The wind momentum and energy deposition associated with the MS evolution is given and the adopted wind properties are discussed. From our atmosphere calculations, which include the effect of multiple scattering and line overlap, we also derive theoretical estimates of mass loss driven by radiation pressure. These values are compared with the predictions from recent wind models of the Munich group (Pauldrach et al. 1990, 1994, Puls et al. 1995). We find an overall agreement with most of their results. In addition, our models are better in reproducing the strong wind momentum rates observed in supergiants than those of Puls et al.

A comparison between boundary conditions given by the conventional plane parallel and the new spherically expanding atmosphere approach is made. For the MS evolution the evolutionary tracks and the interior evolution are found to be basically unchanged by the new treatment of the outer layers. However, for stars close to the Eddington limit, a small uncertainty in the behaviour of the deep atmosphere is found which might marginally affect the evolution. Given the small spherical extension of the continuum forming layers in the considered evolutionary phases, the predicted stellar parameters differ negligibly from those obtained using plane parallel atmospheres.

*Send offprint requests to*: D. Schaerer

## 1. Introduction

This is the first paper in a series dealing with "combined stellar structure and atmosphere models". In the present publication we introduce our method and study wind properties and the interior evolution on the main sequence. The second paper in this series will deal with the spectral evolution of the models presented here (Schaerer et al. 1995, hereafter Paper II). A first study covering the Wolf-Rayet phases is presented in Schaerer (1995a,b). The *CoStar* models use up-to-date input physics and are based on state-of-the-art techniques for both the interior and atmosphere modeling. The latter, in particular, includes the effects of line blanketing in the non–LTE expanding atmosphere, as presented by Schaerer & Schmutz (1994a,b, henceforth SS94ab).

Mass loss through stellar winds is *(i)* the determining process for the evolution of the most massive stars ($M_i \gtrsim 20 M_\odot$; cf. Chiosi & Maeder 1986, Maeder & Conti 1994), and is also responsible for *(ii)* profoundly shaping the emergent spectral energy distribution of these stars (e.g. Kudritzki & Hummer 1990, Schmutz et al. 1992). Since the mass outflow is important in the outermost layers, where radiation decouples from matter, stellar winds cal depth $\tau \leq \tau_\star$ can be derived, in particular also the photospheric radius $R_{\tau=2/3}$ at $\tau_{\rm Ross} = 2/3$. The boundary conditions are given by the velocity

$$v(M_r = M) = v_\star, \qquad (2)$$

which, given the atmospheric structure (see below), is determined by the above requirement for the optical depth. For the interior below $\tau_\star$ expansion is neglected (but see Schaerer 1995a,b for Wolf–Rayet phases). From the continuity equation we obtain the boundary condition for the density $\rho_\star$, i.e.

$$\rho(M_r = M) \equiv \rho_\star = \frac{\dot{M}}{4\pi R_\star^2 v_\star}. \qquad (3)$$

The last boundary condition is given by the temperature $T(R_\star)$. It is obtained by a simultaneous solution of the temperature and density structure for the photosphere and the wind as described in the following.

In calculating the atmospheric structure for the MS phases we closely follow de Koter (1993) and de Koter et al. (1995). Basically the atmosphere is characterised by two parts: the subsonic regime with an extended photosphere, and the wind, where the flow is accelerated to the terminal wind velocity $v_\infty$. In between, both parts are smoothly connected. More precisely, in the subsonic regime the density structure $\rho(r)$ is computed by solving the momentum equation for a stationary flow taking into account gas and radiation pressure. We write the momentum equation as

$$v\frac{\partial v}{\partial r} = -\frac{1}{\rho}\frac{\partial P_g}{\partial r} - \frac{GM\,[1 - \Gamma(r)]}{r^2}, \qquad (4)$$

where

$$\Gamma(r) = \frac{g_R}{g_\star} = \frac{L_\star \kappa_F(r)}{4\pi cGM} = \frac{\sigma T_{\rm eff}^4 \kappa_F(r)}{cg_\star} \qquad (5)$$

is the ratio between the radiative acceleration $g_R$ and the gravitational acceleration $g_\star = GM/R_\star^2$. The other variables have their usual meaning. Note that $\Gamma(r)$ is not constant due to its dependence on the flux weighted opacity $\kappa_F$, which is a priori a function of $T$, $\rho$, $v$ and $dv/dr$ if the effects of continuum and line acceleration are taken into account.

With regard to the flux mean opacity, we have chosen the following approach: For optical depths $\tau_{\rm Ross} < \tau_c = 2$, we adopt the electron scattering opacity, i.e. $\kappa_F = \sigma_e$. Since $\sigma_e$ depends on the chemical composition and the ionization of the considered elements, we adopt the value calculated from the equation of state of the interior model at the boundary $\tau_\star$. At large optical depths ($\tau_{\rm Ross} \geq \tau_c$) we adopt for $\kappa_F(r)$ the Rosseland opacities as for the interior. Note that these opacity calculations assume LTE and a static medium. By checking the departure coefficients in depths LTE is attained to sufficient precision. Since opacities in expanding media are larger than in the static case (e.g. Karp et al. 1977), and the use of Rosseland opacities instead of flux weighted mean opacities can also lead to an underestimate of the opacity, the adopted opacities provide a lower limit. An improved treatment would clearly require a much more complicated calculation of the radiative forces coupled together with a hydrodynamic solution.

Following de Koter et al. (1995) we rewrite Eq. (4) using the continuity equation and write the equation of state for the perfect gas as $P_g = a^2 \rho$, where $a = (kT/\mu m_H)^{1/2}$ is the local isothermal sound speed. This yields

$$v\frac{\partial v}{\partial r} = \frac{1}{v^2 - a^2}\left\{\frac{2a^2}{r} - \frac{GM\,[1 - \Gamma(r)]}{r^2} - \frac{k}{\mu m_H}\frac{\partial T}{\partial r}\right\}. \qquad (6)$$

The mean molecular weight $\mu$ is obtained from the equation of state as described for the electron scattering opacity above. Integrating Eq. (6) from subsonic velocities $v_\star$ outward one progressively obtains an increasing velocity gradient, as $v < a$ and the quantity in large brackets in Eq. (4) is negative for $v < a$. This property allows us to smoothly connect the subsonic part with a wind structure given by

$$v(r) = v_\infty \left(1 - \frac{r_{\rm o}}{r}\right)^\beta, \qquad (7)$$

where the radius $r_{\rm o}$ is adjusted such that both the velocity field and its first derivative are continuous. The choice of the wind parameters $v_\infty$ and $\beta$ is described in Sect. 2.4.

A consistent solution of the momentum equation and temperature structure yields the density and temperature structure. At the same time the velocity $v_\star$ is also adjusted to fulfil the boundary condition with the stellar interior (cf. Eq. (1)). The temperature structure is given by radiative equilibrium in an extended grey atmosphere. It is determined from the generalised Eddington approximation following Lucy (1971) and Wessolowski et al. (1988). The procedure is basically the same as described in SS94a, where more details can be found (cf. also de Koter 1993). In the outermost regions of the wind the temperature is not allowed to drop below a minimum value $T_{\rm min}$, which, following the results of Drew (1989), is chosen as follows: For $T_{\rm eff} \geq 20$ kK we set $T_{\rm min} = 0.4\,T_{\rm eff}$, and $T_{\rm min} = 0.5\,T_{\rm eff}$ for $T_{\rm eff} < 20$ kK.

Because of the large computational effort required, we have neglected the influence of line blanketing on the temperature structure in calculating *CoStar* models. However, the effect of line blanketing is taken into account in the statistical equilibrium and radiative transfer calculations (see Sect. 2.5).

### 2.3. Consistent interior and atmosphere solution

For given stellar and wind parameters ($L_\star$, $R_\star$, $M$ and $\dot{M}$, $v_\infty$, $\beta$) we obtain with the procedure of Sect. 2.2 a consistent atmospheric structure covering the photosphere and ture $T(R_\star)$ determining the temperature boundary condition for the stellar interior. In another iterative process, embracing both the stellar interior and the atmosphere, we finally obtain a consistent solution for the entire star satisfying all boundary conditions. The procedure is schematically shown in Fig. 1.

### 2.4. Mass loss rates and wind structure

The adopted mass loss rate $\dot{M}_{\rm evol}$ and the additional parameters required to describe the wind structure are as follows: [1].

- Mass loss rates are adopted as in Meynet et al. (1994). This means that for population I stars throughout the HR diagram we use the mass loss rates given by de Jager et al. (1988), enhanced by a factor of two. Justifications for this choice are given by Meynet et al. (1994) and Maeder & Meynet (1994).
- The terminal velocities $v_\infty$ are from wind models of Leitherer et al. (1992). Comparisons of our adopted terminal velocities with observations are discussed in Sect. 3.2.1.
- For the rate of acceleration of the supersonic flow (see Eq. (7)), we take $\beta = 0.8$ following theoretical predictions of Friend & Abbott (1986) and Pauldrach et al. (1986). These predictions are in good agreement with observations of O stars by Groenewegen & Lamers (1991).

### 2.5. non–LTE radiation transfer including line blanketing

The non–LTE radiation transfer calculations, which yield the detailed spectral evolution use the atmospheric structure from the *CoStar* model described above. For the detailed transfer calculations we used the *ISA–WIND* non–LTE code of de Koter et al. (1993, 1995). In this code, the line transfer problem is treated using the Sobolev approximation, including the effects of the diffuse radiation field, and the continuous opacity inside the line resonance zone. As a new feature of the *ISA–WIND* code we also include line blanketing, following the opacity sampling technique introduced by Schmutz (1991). The method involves a Monte Carlo radiation transfer calculation including the most important spectral lines of all elements up to zinc. The ionization and excitation of the metals is treated as in SS94ab, where the reader is referred to for a detailed description of the entire procedure.

The input physics for the atmospheric structure calculations consists of atomic data for the elements explicitly included in the non–LTE model. In the present work hydrogen and helium are treated using the same data as de

---

[1] Note that the mass loss rate $\dot{M}_{\rm evol}$ *adopted* for the evolutionary calculations should not be confused with the theoretical mass loss rate estimate $\dot{M}_{\rm calc}$ *predicted* by our atmosphere calculations in Sect. 3.2.2

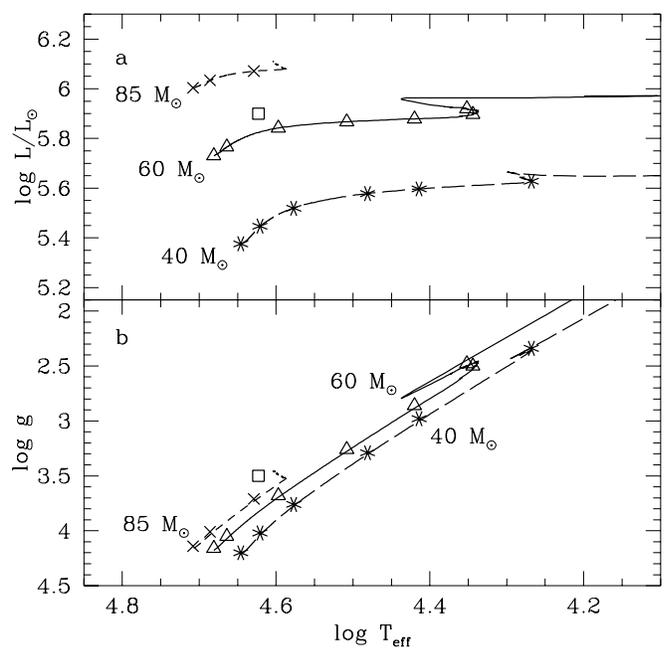

**Fig. 2. a** HR–diagram covering the MS phases for initial masses of 40, 60, and 85 $M_\odot$. The WR stage during the H–burning phase of the 85 $M_\odot$ model is excluded. Crosses, triangles and stars denote the selected models on the 85, 60, and 40 $M_\odot$ tracks respectively, for which detailed spectra have been calculated (see Paper II). The square denotes the position of an additional model presented in Paper II. **b** $\log g$–$\log T_{\rm eff}$ diagram corresponding to the upper panel

Koter et al. (1995): In the atomic model for H and He II we account for the first ten levels with principal quantum number $n = 1$ to 10. The atomic model for He I, consising of 17 levels, is described by Wessolowski et al. (1988). In total we account for 226 line transitions. The inclusion of C, N, O, Si, and other elements is in progress.

The H, He, C, N, and O composition of the atmosphere is that corresponding to the outermost layer of the interior model. For the metals included in the line blanketed atmosphere, the abundances of Anders & Grevesse (1989) have been adopted.

### 3. Interior evolution and feedback to the ISM

We have calculated three *CoStar* tracks for solar metallicity and initial masses of $M_i = 40$, 60, and 85 $M_\odot$. Only the results for the MS phase are discussed in this work. We also exclude those parts of the tracks where the WR phase (defined by a hydrogen surface abundance $X < 0.4$ in mass fraction and $\log T_{\rm eff} > 4.$; cf. Schaller et al. 1992) is already entered during the H–burning phase. In our models this occurs for the 85 $M_\odot$ model. Therefore only part of the H–burning phase of this model is covered in

**Table 1.** H–burning lifetimes and total momentum and energy deposition during MS evolution at Z=0.02

| Initial mass | H–burning phase [$10^6$ yr] | momentum [g cm s$^{-1}$] | energy [erg] | comments |
|---|---|---|---|---|
| 40 $M_\odot$ | 4.40 | $3.12\ 10^{42}$ | $3.36\ 10^{50}$ | |
| 60 $M_\odot$ | 3.47 | $7.87\ 10^{42}$ | $8.25\ 10^{50}$ | |
| 85 $M_\odot$ | 2.67 | $1.42\ 10^{42}$ | $1.75\ 10^{51}$ | to beginning of WR phase |
| | 2.95 | | | total lifetime |

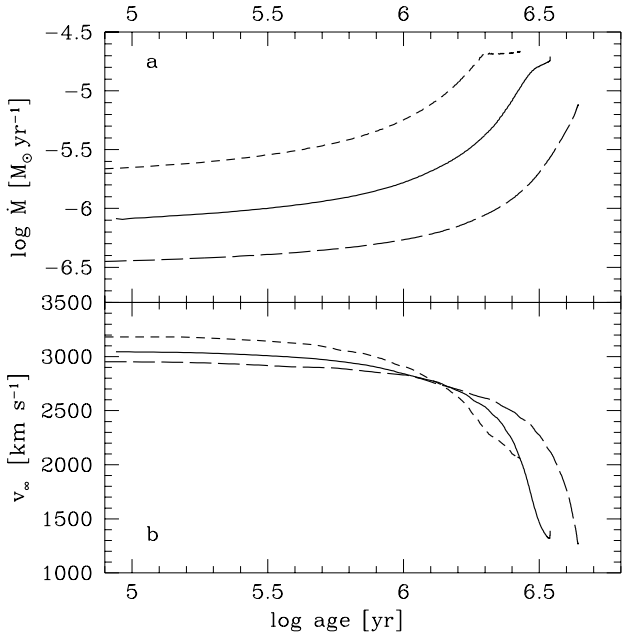

**Fig. 3. a** Mass loss rate as a function of age during the MS evolution. The line styles are as follows: 85 $M_\odot$ short-dashed, 60 $M_\odot$ solid, and 40 $M_\odot$ long-dashed. **b** Terminal velocity as a function of age. The lines are coded as in panel a the present publication. Later phases, including the Wolf–Rayet stages, are discussed in Schaerer (1995a,b).

The evolutionary tracks of the models are presented in Fig. 2. Both the HR–diagram, and the corresponding gravity–effective temperature diagram are shown. The H–burning lifetimes are given in Table 1. As stated above, the $M_i = 85\ M_\odot$ model already enters the H rich WR phase during H–burning. However, for this mass the "normal" O star phase covers 90 % of the total MS phase (cf. Table 1). A comparison of the lifetimes with the results from Meynet et al. (1994) (who use the usual plane parallel grey atmosphere as the outer boundary condition) shows that our 40 and 60 $M_\odot$ models differ by less than 0.2 %. The largest difference, $\sim$ 1.5 %, is obtained for the 85 $M_\odot$ model. This shows that the overall MS evolution for these stars is not affected by the treatment of the atmosphere. In Section 4 we will examine in more details the atmospheric structures to understand this result.

The evolution of the wind properties during the MS evolution are presented in Fig. 3, where we show the mass loss rate, $\dot{M}_{\rm evol}$, and the terminal velocity, $v_\infty$, as a function of age. Due to an increasing luminosity and evolution towards lower temperatures $\dot{M}_{\rm evol}$ increases during the evolution. The terminal velocity remains nearly constant during the first million years, whereas it decreases afterwards, mainly as a result of an increasing radius, hence a decreasing escape velocity.

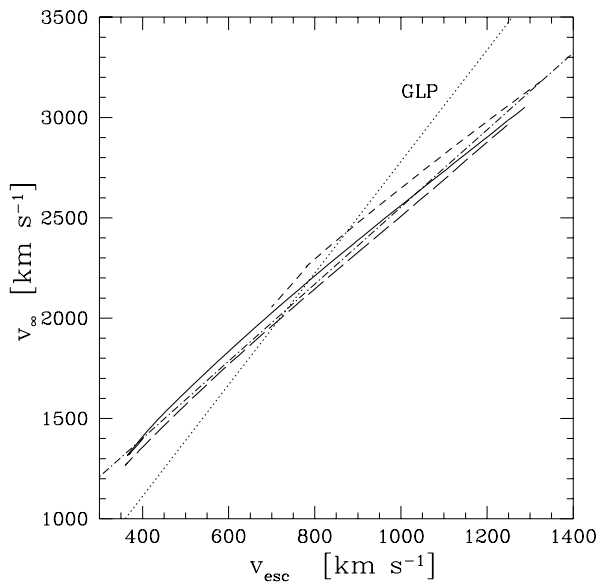

**Fig. 5.** Adopted terminal velocities as a function of escape velocity (The thick lines denote our models: line styles as in Fig. 3). Thin lines represent different fits. The thin dashed-dotted line shows the fit through our model data (see text). The dotted line (marked GLP) is the relation obtained by Groenewegen et al. (1989) from UV fits. See text for a discussion

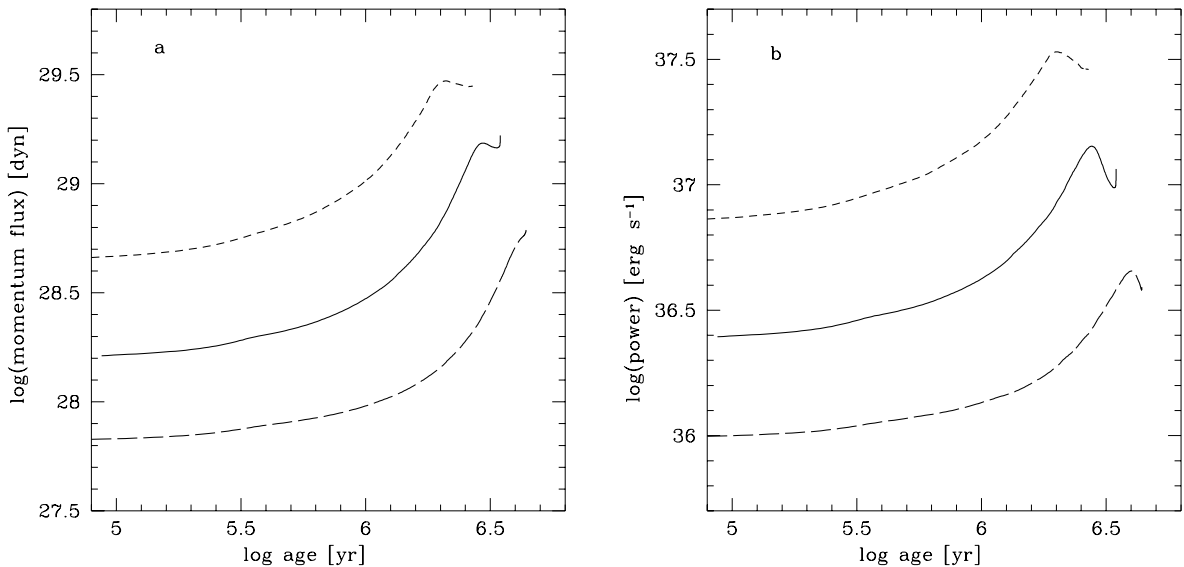

**Fig. 4. a** Momentum flux of the 85 (short-dashed), 60 (solid), and 40 $M_\odot$ (long-dashed) model as a function of its age. **b** Mechanical wind power emitted as a function of the age. The lines are coded as in the panel a

## 3.1. Deposition of momentum and energy

To study the effect of mass loss in massive stars on the interstellar medium it is of interest to quantify the wind momentum flux $\dot{M}_{\rm evol}v_\infty$, and the rate of release of mechanical energy $1/2\,\dot{M}_{\rm evol}v_\infty^2$ into their surroundings. These values may be used to investigate the properties of individual nebulæ, but may also be used to derive predictions for integrated young stellar populations (see Leitherer et al. 1992, Williams & Perry 1994)

The momentum flux and mechanical wind power during the MS phase is shown in Fig. 4, while the total momentum and energy deposition integrated over the MS lifetime is given in Table 1. Figure 4 illustrates the progressive increase of both wind momentum and energy during the MS evolution caused by the strong enhancement of mass loss towards the end of the main sequence.

## 3.2. Discussion of wind properties

In this section, we discuss the most important *adopted* wind parameters, i.e. terminal velocity and mass loss rate. We first compare our adopted $v_\infty$ with observations. We then present theoretical mass loss rates derived from simple energetic considerations, and compare these with the adopted values.

### 3.2.1. Terminal velocities

To make a comparison between the adopted and observed values of $v_\infty$, we plot both against the escape velocity $v_{\rm esc}$. The relation between $v_{\rm esc}$ and $v_\infty$ is discussed by e.g.

Castor et al. (1975) and Abbott (1978). We derive the escape velocity from the effective gravity $g_{\rm eff} = g_\star(1-\Gamma)$. The correction for the radiation pressure due to electron scattering is given by $\Gamma = 7.66\,10^{-5}\sigma_e\,(L/L_\odot)\,(M_\odot/M)$, where the value of the electron scattering opacity corresponds to the value at the boundary $R_\star$ (cf. Sect. 2.2).

In Fig. 5 the adopted terminal velocity is shown as a function of the resulting $v_{\rm esc}$. A least-square fit to the values from the three tracks shown in Fig. 5 yields $v_\infty = (633.97 \pm 4.42) + (1.921 \pm 0.005)\,v_{\rm esc}$, where the velocities are in km s$^{-1}$. As a comparison we have also plotted in Fig. 5 the relation derived by Groenewegen et al. (1989, Eq. 6) from UV fits including the effect of turbulence. The agreement is reasonable, the maximum differences in the considered range being about 20 %. Note that the overestimation obtained by their Eq. 6 for stars close to the ZAMS (i.e. large $v_{\rm esc}$) would be slightly reduced by adopting their fit-formula (Eq. 9), which takes also the dependence of $v_\infty$ on $T_{\rm eff}$ into account. For velocities $v_\infty \lesssim 2000$ km s$^{-1}$ our adopted formula probably slightly overestimates the wind velocity (cf. Leitherer et al. 1992). Our results are also in agreement with the relation from Prinja et al. (1990), provided a small ($\sim 8$ %) downward correction of their adopted escape velocities is applied.

In résumé, we can conclude that for escape velocities $v_{\rm esc} \gtrsim 800$ km s$^{-1}$ the adopted terminal velocities compare well with the observations, while for lower values the adopted $v_\infty$ is probably overestimated by up to $\sim 25$ %. However, as can be seen from Fig. 3, this only concerns a short period of time close to the end of the main sequence evolution.

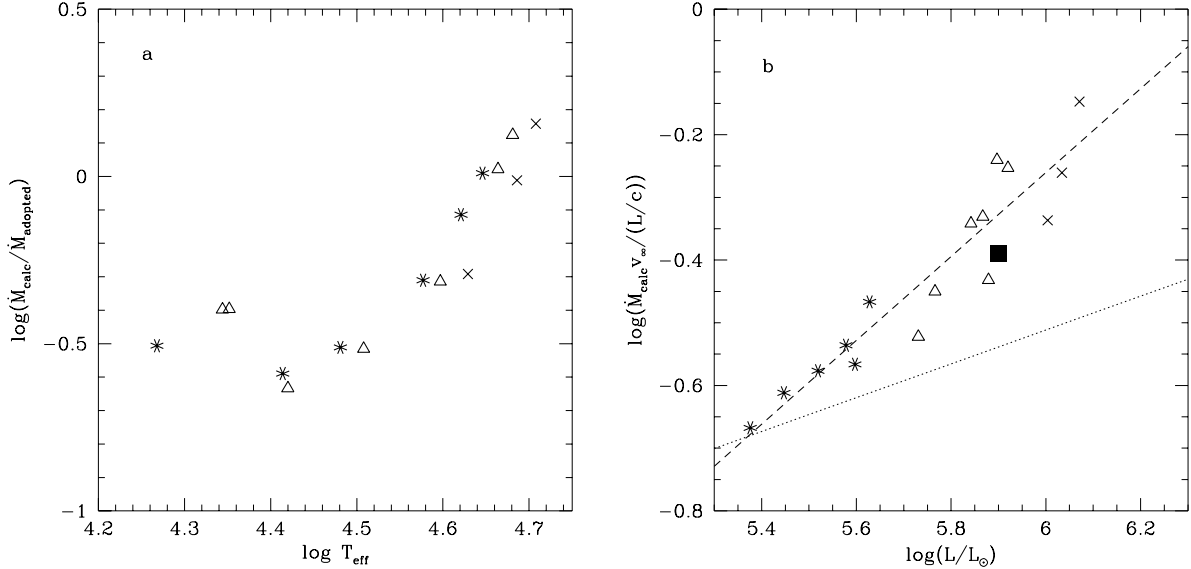

**Fig. 6. a** Logarithm of the ratio of the theoretical mass loss rate $\dot{M}_{\rm calc}$ to the adopted value $\dot{M}$ as a function of effective temperature. Crosses, triangles and stars denote the selected models on the 85, 60, and 40 $M_\odot$ tracks respectively. For a discussion see Sect. 3.2.2. **b** Theoretical wind momentum efficiency $\eta$ as a function of luminosity (same symbols as in panel (a)). Dashed line: Fit to the model predictions (Eq. (8)). The filled square indicates the result obtained from a consistent hydrodynamic model for $\zeta$ Puppis (SS94a, model A). Dotted lined: Wind efficiency predicted by Lamers & Leitherer (1993, Eq. 23). For our main sequence models the predicted wind efficiencies are between $\sim 0.2$ and $0.7$, which is up to a factor of 2 larger than the values calculated by Lamers & Leitherer using the line force parameters of Pauldrach et al. (1990)

### 3.2.2. A mass loss rate estimate using the photon energy balance

Presently all evolutionary models, including the ones developed in this work, rely on the use of empirical mass loss rates. While radiation driven wind theory is quite successful in explaining the overall properties of OB stars and possibly also LBV's (e.g. Kudritzki et al. 1991 and references therein; Pauldrach & Puls 1990) discrepancies still remain (e.g. Groenewegen et al. 1989, Schmutz & Schaerer 1992, Lamers & Leitherer 1993 hereafter LL93, Puls et al. 1995). Although the calculation of consistent hydrodynamic wind models, such as the ones presented by SS94ab, is clearly beyond the scope of the present work, it is however very interesting to make estimates of the mass loss rates which can be driven by radiation pressure.

We estimate the mass loss rate, subsequently called $\dot{M}_{\rm calc}$, from the photon energy balance, which is obtained from our Monte-Carlo radiation transfer calculations. Since our models cover the entire main sequence for stars from $M_i = 40$ to $85\ M_\odot$, a comparison of estimated radiation driven mass loss rates with observations could yield useful insight to understand the present difficulties of the radiation driven wind theory.

To determine $\dot{M}_{\rm calc}$, we follow the considerations of Abbott & Lucy (1985). From the total radiative energy deposition in the wind, $L_T = L(R_\star) - L(\infty)$, we calculate $\dot{M}_{\rm calc}$ from $L_T = 1/2\ \dot{M}_{\rm calc}[v_\infty^2 + v_{\rm esc}^2]$, assuming that the entire radiative energy deposition in the wind is converted to mechanical energy only, by lifting matter out of the gravitational field and giving it its asymptotic kinetic energy. The flux transfer rate $L_T$ is obtained from the MC simulation taking into account a large number of metal lines, and including line overlap and multiple scattering (see SS94a and Paper II). Note, however, that achieving consistency is beyond the scope of the present work. This could be accomplished by adjusting the mass loss rate until $L_T$, which depends on the wind density, equals the mechanical energy (cf. Abbott & Lucy 1985). Our results should only therefore be taken as estimates for mass loss driven by radiation pressure. Surprisingly, as shown below, this method shows a good agreement with detailed hydrodynamic calculations of Schaerer & Schmutz (1994a).

In Figure 6a, we have plotted the ratio $\dot{M}_{\rm calc}/\dot{M}_{\rm evol}$ of the theoretical to the adopted mass loss rate as a function of effective temperature. For the models close to the ZAMS ($T_{\rm eff} \gtrsim 43.6$ kK) we see that the estimated theoretical mass loss rate is of the same order, or even larger, than the adopted values for $\dot{M}_{\rm evol}$. In this case, the energy extracted from the radiation field can, in principle, account for the energy stored in the stellar wind. On the other hand, for the models with $T_{\rm eff} \lesssim 43.6$ kK, the adopted $\dot{M}_{\rm evol}$ is larger than the theoretical derived values. Similar discrepancies as a function of effective temperature are predicted by LL93, who use analytic solutions of Ku-

Pauldrach et al. (1990). This indicates a shortcoming in the current state of the radiation driven wind theory for evolved stars, which may be resolved by consistent hydrodynamic calculations.

Of particular interest is the wind efficiency $\eta = \dot{M} v_\infty / (L_\star/c)$ predicted by radiation driven wind models. For OB stars one typically finds $\eta \sim 0.1$–$0.6$, while for WR stars values considerably larger than unity can be obtained (e.g. LL93).

In Figure 6b, we have plotted the theoretical wind efficiency $\eta_{\rm calc}$ as a function of the luminosity. The model data are fitted by the relation

$$\log \eta_{\rm calc} = (0.669 \pm 0.067) \log(L/L_\odot) - 4.276 \pm 0.386, \quad (8)$$

with a rms of 0.052 dex. The relation is represented by the dashed line. Interestingly the corresponding value from a self-consistent hydrodynamic calculation of SS94a for $\zeta$ Puppis is quite well matched by the above relation. This indicates that, despite the lack of consistency pointed out above, the estimated mass loss rate $\dot{M}_{\rm calc}$ determined from the flux transfer rate $L_T$ yields a reasonable value for the mass loss rate as derived from full hydrodynamic modeling.

We have compared our predictions for $\eta_{\rm calc}$ with the those of LL93, which are based on parametrised line forces calculated by Pauldrach et al. (1990). The results of LL93 (their Eq. 23) are plotted as the dotted line in Fig. 6b, the uncertainty being 0.1 dex. Figure 6b reveals that our models predict a steeper increase of the wind momentum with luminosity. For the most luminous model we obtain $\eta_{\rm calc} \sim 0.7$, which is a factor of two larger than the value of LL93. Since the methods used in both approaches are completely independent, it is difficult to trace the differences back to one single reason. However, we expect to find the largest differences for cases where the effects of multiple scattering may become important since this effect is correctly treated in our models, but has been neglected by Pauldrach et al. (1990). On the other hand the larger wind efficiencies predicted by our calculations could also indicate systematic differences in the ionization structures. With respect to the ionization problem, we note that our models are supported by comparisons of predicted Fe features in UV spectra, which show a good agreement with observations for the evolved models (see Paper II), where iron is the dominant source of the radiation force.

New calculations of radiation driven wind models based on the improvements reported by Pauldrach et al. (1994) have recently been presented by Puls et al. (1995). They show that the most appropriate quantity when comparing theoretical predictions with observations is the "wind momentum rate" $\dot{M} v_\infty R_\star^{1/2}$, since this quantity is expected to show a very weak dependence on the adopted stellar parameters. Indeed a strong correlation of $\log \left( \dot{M} v_\infty R_\star^{1/2} \right)$ with $\log L$ is expected if the line force

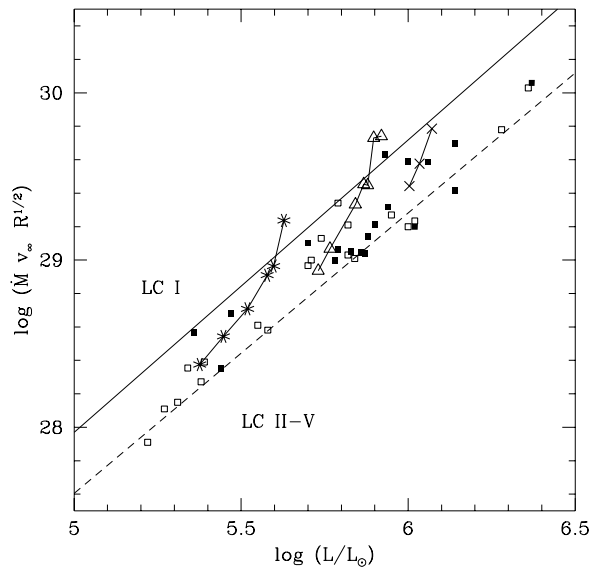

**Fig. 7.** Theoretical wind momentum rate as a function of luminosity. Stars, triangles and crosses denote the predictions from our models on the 40, 60, and 85 $M_\odot$ tracks respectively. Small squares show the predictions for individual galactic objects of luminosity class I (filled squares) and II-V (open squares) from Puls et al. (1995). The lines show the mean relations for observed Luminosity class I (solid) and II-V objects (dashed) derived from the data of Puls et al. (their Table 8, galactic objects only)

parameter $k$, which represents the flux-weighted number of driving lines, is constant. The slope is then only determined by the line force parameters $\alpha$ and $\delta$, and is found to be $1/(\alpha - \delta)$. Observational evidence for such a correlation has been presented by Kudritzki et al. (1995).

To compare our calculations with the results of Puls et al. (1995) we have plotted the predicted wind momentum rate from our models as a function of luminosity in Fig. 7. Also shown are the predictions for the individual galactic objects from Puls et al. and the mean observed values for luminosity class I objects (solid) and the objects of class II-V (dashed). Figure 7 shows that our models reproduce well the observed behaviour, i.e. both the slope of the relation and the trend of increased wind momentum rate for supergiants. On the average our predictions seem to be in rough agreement with the results of Puls et al. (1995). However, as pointed out by Puls et al. and apparent in Fig. 7, their models show a discrepancy for the supergiants as their theoretically predicted momentum is about 0.25 dex too small. This discrepancy is not present in our models, hence our supergiant models are in better agreement with observations. Whether this difference with the recent Munich models is due to the effect of multiple scattering, which in contrast to Puls et al. is included in our calculations, or has other causes (different careful investigation. As mentioned above iron, which is the dominant source of the radiation force, is, however, well described in our supergiant models (see Paper II).

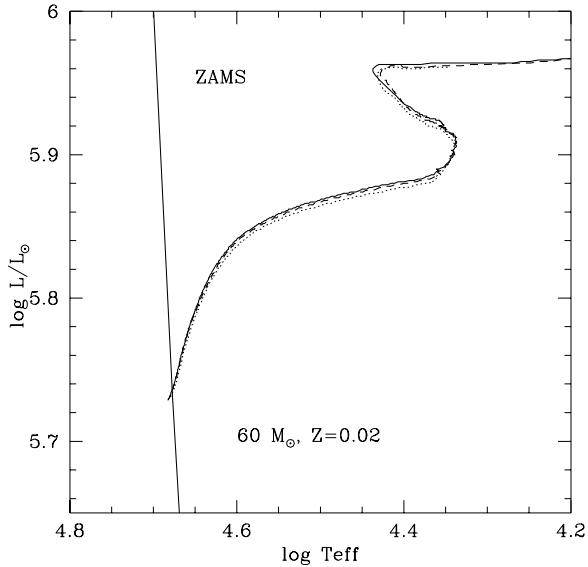

**Fig. 8.** HR–diagram for the initially 60 $M_\odot$ model, calculated with *CoStar* (solid line) and the standard Geneva evolution code (dashed). In both cases the boundary with the stellar interior model is set at $\tau_{\rm Ross}$=20. The standard track (plane parallel atmosphere, $\tau_{\rm Ross} = 2/3$) is shown with the dotted line. Also plotted is the zero-age main sequence (ZAMS). The difference between the tracks is negligible in view of the errors related to observational determinations of temperatures and luminosities

## 4. Comparison of plane parallel and extended atmospheres for evolutionary models

In this section we discuss the influence of atmospheric boundary conditions on the evolutionary tracks. As will be discussed below, we need to adopt a slightly different treatment for the most massive stars ($M_i \gtrsim 85 M_\odot$), which lie close to the Eddington limit. We therefore discuss these cases separately.

### 4.1. Evolution of the $M_i = 40$ & $60$ $M_\odot$ models

In Fig. 8 we show a detailed comparison between the 60 $M_\odot$ track calculated with the conventional atmospheric treatment and the stellar models including spherically expanding atmosphere (*CoStar* models). To compare exactly the same definitions of effective temperatures, and hence radii, we have set the boundary at $\tau_{\rm Ross}$=20 for both cases. Different tracks should thus reflect the influence of different boundary conditions between plane parallel lated with the standard treatment, i.e. plane parallel atmosphere with the inner boundary at $\tau_{\rm Ross} = 2/3$ (dotted line).

As we can see from Fig. 8 the tracks hardly differ. Similarly, the internal structures are basically identical, which explains why the H–burning lifetimes do not differ. This is due to the fact that the wind is mostly optically thin, which implies that the thermal conditions are essentially determined in the quasi-hydrostatic photosphere and hence the boundary conditions including the wind are nearly identical to those of plane parallel models.

With respect to the standard track, both the plane parallel and spherically expanding models with the inner boundary set at large optical depth ($\tau_{\rm Ross}$=20) evolve at slightly higher temperature during most of the MS. The difference amounts to only $\Delta T \lesssim 400$ K, if the same ages are compared. On the other hand the spherical extension, measured by the ratio of radius $R_{2/3}$ at $\tau_{\rm Ross} = 2/3$ to $R_\star$ reaches only up to $R_{2/3}/R_\star \lesssim 1.015$ for the most evolved models. This confirms the expectation that despite the spherical extension of the atmosphere the radius definition causes no ambiguity for the MS phase of stars with $M_i = 40$ & 60 $M_\odot$.

### 4.2. Evolution of the $M_i = 85$ $M_\odot$ model

For more massive stars the situation can become slightly more complicated because they evolve quite close to the Eddington limit. This may introduce an additional difficulty in the computation of the atmospheric structure.

To illustrate this point we have calculated the MS evolution of a 85 $M_\odot$ model with both types of atmospheres. We first assumed a constant mass loss rate of $\dot{M}_{\rm evol}$=1.14 $10^{-5}$ $M_\odot\,{\rm yr}^{-1}$ to eliminate indirect effects of mass loss on the evolutionary tracks (see Sect. 4.2.2). The resulting tracks are shown in Fig. 9a. It is important to note that, contrary to the usual treatment (cf. Sect. 2.2), the atmosphere of the *CoStar* model has been calculated with a depth-independent effective gravity corrected for electron scattering only. As we will argue below, this should yield a more realistic structure than the hydrostatic plane parallel atmosphere using Rosseland opacities.

#### 4.2.1. Constant mass loss tracks for $M_i = 85$ $M_\odot$

Figure 9b shows the evolutionary track in the gravity–$T_{\rm eff}$ diagram. To illustrate its proximity with respect to the Eddington limit we have plotted the limit derived by Lamers & Fitzpatrick (1988) from Kurucz models (labeled $g_{\rm Edd}$). Also shown is the lowest $\log g$ value (labeled $\log g_{\rm min}$) for which they obtained converged hydrostatic atmosphere structures, and which were used to determine the $g_{\rm Edd}$ limit by extrapolation.

The difference between the tracks in the HR–diagram (Fig. 9a) can be understood by looking at the atmospheric

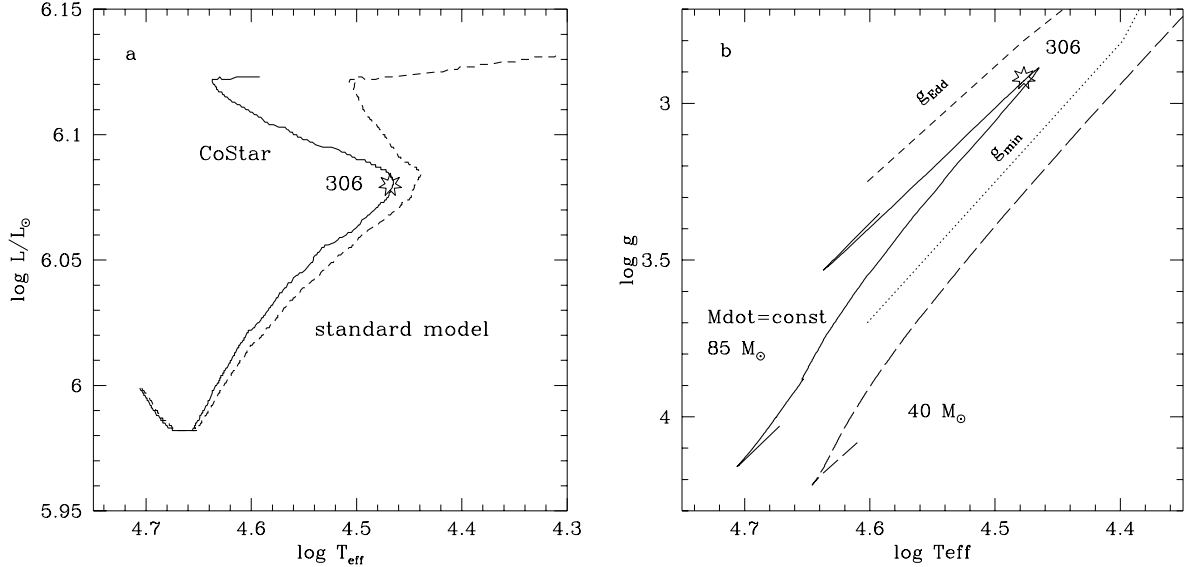

**Fig. 9. a** HR–diagram of a 85 $M_\odot$ model calculated with constant $\dot{M}_{\rm evol}=1.14\,10^{-5}\,M_\odot\,{\rm yr}^{-1}$. The solid line shows the *CoStar* model adopting a depth-independent effective gravity. The dashed track is for the conventional plane parallel atmosphere. The *CoStar* track evolves at slightly higher $T_{\rm eff}$ due to a larger pressure at the boundary $\tau_\star$. The star marks a model discussed in detail (cf. Fig. 10). **b** $\log g$ vs. $\log T_{\rm eff}$ diagram for the 85 $M_\odot$ track calculated with constant $\dot{M}_{\rm evol}$ (solid line). The star marks the position of the same model as in panel (a). As a comparison the track of the 40 $M_\odot$ model from Fig. 2 is also plotted (long dashed). The 60 $M_\odot$ track is not shown here, since it virtually coincides with the dotted line. The short dashed line, marked $g_{\rm Edd}$, traces the Eddington limit determined by Lamers & Fitzpatrick (1988) from Kurucz models by extrapolation from the lowest gravity models shown by the dotted line ($g_{\rm min}$). This illustrates the proximity of the 85 $M_\odot$ track to the Eddington limit

structures of models with stellar parameters corresponding to the model marked 306 in this figure. Figure 10 shows the temperature and density stratification of the plane parallel (dashed line) and the spherically expanding atmosphere (solid line). Note that at $\log \tau_{\rm Ross} \gtrsim -0.2$ the hydrostatic model shows a small density inversion. The $\rho$–inversion occurs in the zone where He becomes completely ionised and is due to the increasing opacity (Fig. 9b).

If we suppress the density inversion in the plane parallel model by assuming a constant opacity ($\kappa = \sigma_e$), the density scale height remains constant (dotted line). However, a higher density results at any given optical depth, and at the inner boundary in particular, because the opacity only accounts for electron scattering.

In a hydrodynamic solution, on the other hand, one could make the conjecture that the opacity increase at rather low optical depths contributes to the acceleration of the outflow and washes out the density inversion. Since in this work we are not able to solve consistently for the full hydrodynamic equations of the entire atmospheric structure, we use our usual procedure to describe the wind and the photosphere, but we do not allow for density inversions in *CoStar* models. This is simply obtained by adopting a constant, i.e. depth-independent effective gravity corrected for electron scattering only. The usual Rosseland opacities are however used for the temperature determination. The resulting structure of the spherically expanding atmosphere is plotted in Fig. 10 showing (in the inner part) the same density scale height as the plane parallel model where the $\rho$–inversion has been suppressed. One can clearly see the large velocity gradient, which is just located outward of the zone of the density inversion of the hydrostatic model.

Since the temperature is basically determined in the photosphere (cf. above) both models have the same temperature at the inner boundary. Due to the higher density in atmospheres without $\rho$–inversion the pressure is, however, slightly larger [2] than in the hydrostatic atmosphere where the density inversion occurs.

When the modified boundary conditions are used to determine the entire stellar structure, the star will readjust itself to the increased compression, which yields a slightly lower radius. This explains the differences between the *CoStar* and the conventional track (both with $\dot{M}_{\rm evol}$=const) shown in Fig. 9. As we will see below, the effect of modified boundary conditions has indirect consequences on the evolution, since the position in the HR–diagram determines the mass loss rate.

---

[2] Approximately 30 % of the total pressure at $\tau_\star$ is provided by the gas pressure.

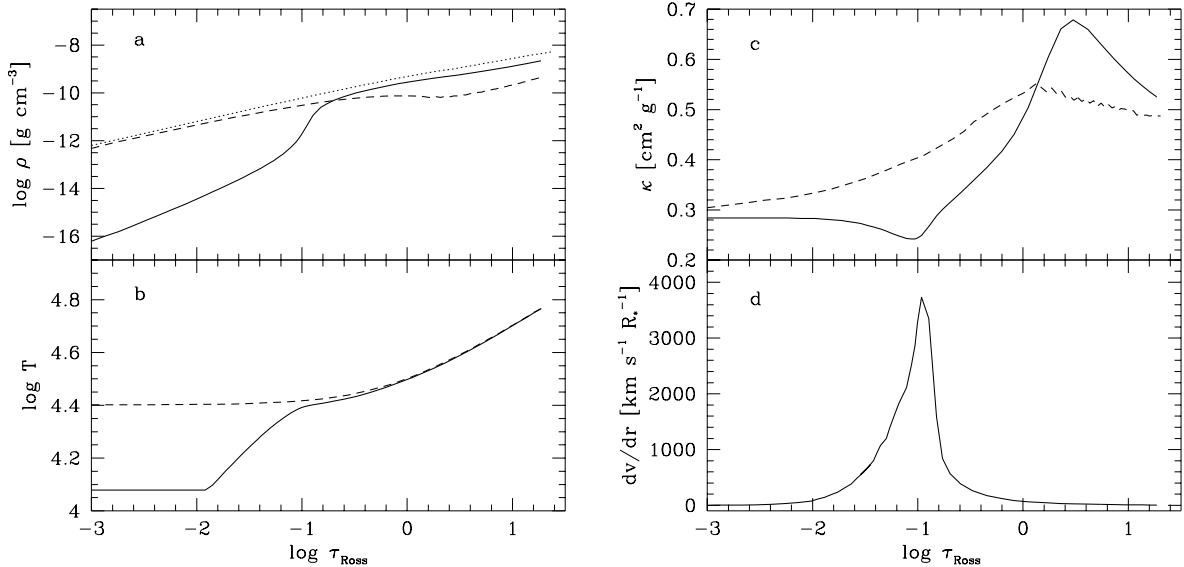

**Fig. 10.** Atmosphere structure for model 306 (cf. Fig. 9). Dashed line: plane parallel atmosphere; dotted line: also plane parallel, but adopting a depth-independent $g_{\rm eff}$ using $\kappa = \sigma_e$; solid line: spherically expanding atmosphere with constant $g_{\rm eff}$ in the photosphere, and $\dot{M}_{\rm evol} = 1.14\,10^{-5}\,{\rm M}_\odot {\rm yr}^{-1}$. **a** Density structure as a function of Rosseland optical depth. **b** Temperature structure. **c** Rosseland opacity as a function of Rosseland optical depth. **d** Velocity gradient of the spherically expanding atmosphere as a function of optical depth.

### 4.2.2. Variable mass loss track for $M_i = 85\,M_\odot$

We now adopt the mass loss rate prescription given in Sect. 2.4. This allows us to illustrate indirect effects of the boundary conditions on evolutionary tracks. Figure 11 shows a comparison of the *CoStar* track with the track calculated with the plane parallel atmosphere (from Meynet et al. 1994). Here we concentrate on the evolution from the ZAMS to the beginning of the WR phase, marked by a star in the HR–diagram. To illustrate the differences in later phases, we show the continuation of the tracks in the WNL phase. As explained above, the *CoStar* track has been calculated with a constant effective gravity in the photosphere. Therefore the pressure at the boundary $\tau_\star$ is slightly larger than for the plane parallel atmosphere, which results in a reduced radius, i.e. a larger effective temperature. Since, in turn, the mass loss rate at a given luminosity decreases with increasing $T_{\rm eff}$ (see de Jager et al. 1988; also Lamers & Leitherer 1993), the *CoStar* model suffers a smaller mass loss (cf. Fig. 11b), which explains its slight overluminosity and hence the shortened lifetime (although only $\sim 1.5\%$) with respect to the conventional model.

Clearly the effects discussed here for the O star phase are quite small. Looking at the beginning of the subsequent WR phase, part of which is also plotted in Fig. 11, the tracks seem to diverge. In fact, this only concerns a very short time before the end of H-burning and the beginning of the hydrogen free WR phase (WNE).

We conclude that the differences in evolutionary track, interior evolution and lifetimes on the main sequence, between the conventional and *CoStar* models are negligible. However, a small uncertainty in the tracks and the predicted stellar parameters remains for the most luminous MS stars due to their proximity to the Eddington limit. A fully hydrodynamic treatment including the subphotospheric layers and the wind is required to improve the present treatment.

Post-MS phases are discussed in Schaerer (1995a,b).

## 5. Summary and conclusions

In the present paper we have presented the first "combined stellar structure and atmosphere models" (*CoStar*) for massive stars, which consistently treat the stellar interior and a spherically expanding atmosphere including the wind. Our approach replaces the widely used boundary conditions given by plane parallel grey atmospheres. The *CoStar* models also predict the detailed emergent spectrum along the evolutionary tracks taking non–LTE effects and line blanketing into account (see Paper II).

As a first exploration of the behaviour of radiation driven wind models on the entire MS we derive estimates of theoretical mass loss rates from energy considerations (Sect. 3.2.2). This is of particular interest since our atmosphere calculations include the effects of multiple scattering and line overlap, which are usually neglected. We have compared our results with the predic-tions from recent wind models of the Munich group (Pauldrach et al. 1990, 1994, Puls et al. 1995). While we find an overall agreement with their results, our models in addition also reproduce the strong wind momentum rates observed in supergiants. Possible reasons for this finding have been discussed. Consistent hydrodynamic calculations (cf. Schaerer & Schmutz 1994a) will be necessary to verify whether for OB stars the quantitative discrepancies of the wind theory (see e.g. Puls et al. 1995) can indeed be explained by the effects included in our models.

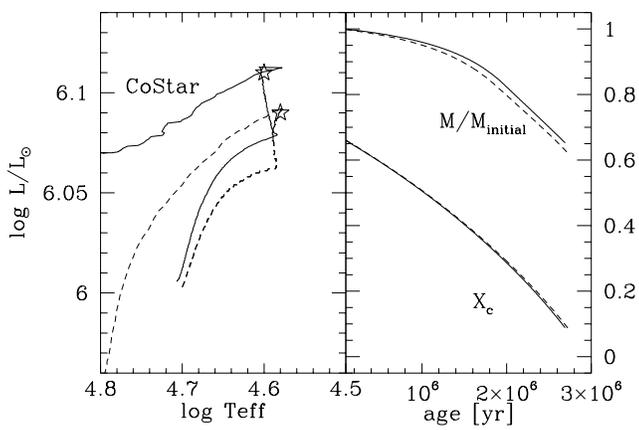

**Fig. 11. a** HR–diagram of the 85 $M_\odot$ *CoStar* model (solid line) and the conventional model from Meynet et al. (1994; dashed line: uncorrected temperature $T_\star$). The H–burning phase proceeds from the ZAMS up to the position marked with a star where the tracks enter the WR phase (WNL). Since the *CoStar* model evolves at slightly higher temperatures its mass loss is smaller (cf. panel (b)) and hence it evolves at higher luminosities. *(b):* Evolution of the central hydrogen abundance $X_c$ (mass fraction) and the stellar mass, expressed by the ratio $M/M_i$ of the present mass to the initial mass. Shown is only the evolution from the ZAMS up to the beginning of the WR phase, which is marked by a star in the panel (a). The mass difference at this stage is $\approx 2.8\ M_\odot$. Although the tracks diverge during the WR phase, the resulting differences for the H–burning lifetimes are still small

One of the important aims of this study was to investigate the possible influence of the spherically extended atmosphere on the main sequence evolution of massive stars, and in particular, on the predicted positions in the HR-diagram. From the well known convergence properties of radiative envelopes (e.g. Schwarzschild 1958) it is expected that small changes of the external boundary conditions have little influence on the interior evolution during phases where only a central nuclear burning source is present. This is confirmed in general by our modeling of the MS evolution, although some uncertainties, related to the modeling of either density inversions in plane parallel hydrostatic atmospheres or the acceleration zone (Sect. 4).

It is important to realize that the above stated relatively unimportant effect of outer boundary conditions on the stellar structure may not generally be true in phases where a burning shell is present. In this case the boundary values may indeed influence the conditions in the shell, leading to a readjustment of its nuclear energy production, which thereby also affects the growth of the underlying He-core. Such a behaviour was e.g. found for stars with $M \sim 15$–$20\ M_\odot$ undergoing blue loops (Langer 1991). Subsequent to the MS evolution presented in this paper the considered models will evolve through a brief "LBV like" phase with strong mass loss before entering the Wolf–Rayet phase (see Meynet et al. 1994), where the burning shell will be extinguished in most cases. While the effects of the expanding atmosphere on the interior evolution has been studied for the WR phases by Schaerer (1995a,b), its influence during the short transient phase is expected to be negligible, although it still needs to be explored.

With respect to the accuracy of positions in the HR-diagram the situation can be summarised as follows: Since for main sequence OB stars the continuum is essentially formed in a quasi-hydrostatic photosphere, not only the subsonic lower boundary conditions of both spherical and plane parallel atmospheres are basically identical, but also the spherical extension is negligible. The predicted stellar parameters (radius, $T_{\rm eff}$) are therefore essentially identical (Sect. 4).

*Acknowledgements.* We thank Dr. Joachim Puls for providing us with results from recent hydrodynamic calculations. This work was supported in part by the Swiss National Foundation of Scientific Research and by NASA through a grant to the GHRS science team.

## References

Abbott D.C., 1978, ApJ 225, 893
Abbott D.C., Lucy L., 1985, ApJ 288, 679
Anders E., Grevesse N., 1989, Geochim. Cosmochim. Acta 53, 197
Castor J.I., Abbott D.C., Klein R.I., 1975, ApJ 195, 157
Chiosi C., Maeder A., 1986, ARA&A 24, 329
de Koter A., 1993, Ph.D. thesis, Utrecht University, The Netherlands
de Koter A., Lamers H.J.G.L.M., Schmutz W., 1995, A&A, in press
de Koter A., Schmutz W., Lamers H.J.G.L.M., 1993, A&A 277, 561
Drew J.E., 1989, ApJS 71, 267
Friend D.B., Abbott D.C., 1986, ApJ 311, 701
Groenewegen M.A.T., Lamers H.J.G.L.M., 1991, A&AS 88, 625
Groenewegen M.A.T., Lamers H.J.G.L.M., Pauldrach A.W.A., 1989, A&A 221, 78
Iglesias C.A., Rogers F.J., Wilson B.G., 1992, ApJ 397, 717
de Jager C., Nieuwenhuijzen H., van der Hucht K.A., 1988, A&A 173, 293

Kudritzki R.P., Gabler R., Kunze D., Pauldrach A., Puls J., 1991, in "Massive Stars in Starbursts", STScI Symp. Series 5, Eds. C. Leitherer, N. Walborn, T.M. Heckman, C.A. Norman, Cambridge Univ. Press, p. 59

Kudritzki R.P., Hummer D.G., 1990, ARA&A 28, 303

Kudritzki R.P., Lennon D.J., Puls J., 1995, in "Science with the Very Large Telescope", Eds. J.R. Walsh, I.J. Danziger, ESO Garching, in press

Kudritzki R.P., Pauldrach A., Puls J., Abbott D.C., 1989, A&A 219, 205

Kurucz R.L., 1991, in "Stellar Atmospheres: Beyond Classical Models", NATO ASI Series C, Vol. 341, Eds. L.Crivellari, I.Hubeny, D.G.Hummer, p. 441

Lamers H.J.G.L.M., Fitzpatrick E.L., 1988 ,ApJ 324, 379

Lamers H.J.G.L.M., Leitherer C., 1993, ApJ 412, 771 (LL93)

Langer N., 1991, A&A 252, 669

Leitherer C., Robert C., Drissen L., 1992, ApJ 401, 596

Lucy L.B., 1971, ApJ 163, 95

Maeder A., 1990, A&AS 84, 139

Maeder A., Conti P., 1994, ARA&A 32, 227

Maeder A., Meynet G., 1994, A&A , 287, 803

Meynet G., Maeder A., Schaller G., Schaerer D., Charbonnel C., 1994, A&AS 103, 97

Pauldrach A., Puls J., 1990, A&A 237, 409

Pauldrach A., Kudritzki R.P., Puls J., Butler K., 1990, A&A 228, 125

Pauldrach A., Kudritzki R.P., Puls J., Butler K., Hunsinger J., 1994, A&A 283, 525

Pauldrach A., Puls J., Kudritzki R.P., 1986, A&A 164, 86

Prinja R.K., Barlow M.J., Howarth I.D., 1990, ApJ 361, 607

Puls J., Kudritzki R.P., Herrero A., Pauldrach A.W.A., Haser S.M., Lennon D.J., Gabler R., Voels S.A., Vilchez J.M., Wachter S., Feldmeier A., 1995, A&A , submitted

Schaerer D., 1995a, Ph.D. thesis No. 2738, Geneva University, Switzerland

Schaerer D., 1995b, A&A , in press

Schaerer D., de Koter A., Schmutz W., 1995, A&A , in press (Paper II)

Schaerer D., Schmutz W., 1994a, A&A 288, 231 (SS94a)

Schaerer D., Schmutz W., 1994b, Space Sci. Rev. 66, 173 (SS94b)

Schaller G., Schaerer D., Meynet G., Maeder A., 1992, A&AS 96, 269

Schmutz W., 1991, in "Stellar Atmospheres: Beyond Classical Models", Eds. Crivellari, L., Hubeny, I., Hummer, D.G., NATO ASI Series C, Vol. 341, p. 191

Schmutz W., Leitherer C., Gruenwald R., 1992, PASP 104, 1164

Schwarzschild M., 1985, Structure and Evolution of the Stars, Dover Publ., New York

Williams R.J.R., Perry J.J., 1994, MNRAS 269, 530

Wessolowski U., Schmutz W., Hamann W.R., 1988, A&A 194, 160